\documentclass[11pt]{elsarticle}

\pdfoutput=1

\makeatletter
\def\ps@pprintTitle{%
  \let\@oddhead\@empty
  \let\@evenhead\@empty
  \let\@oddfoot\@empty
  \let\@evenfoot\@oddfoot
}
\makeatother

\usepackage{url}
\usepackage{breakurl}
\usepackage[breaklinks,
            colorlinks = true,
            linkcolor = blue,
            urlcolor  = blue,
            citecolor = blue,
            anchorcolor = blue]{hyperref}

\usepackage{lineno}

\usepackage{hyperref}

\emergencystretch=\maxdimen
\usepackage[utf8]{inputenc} 
\usepackage[T1]{fontenc} 
\usepackage{comment}
\usepackage{textcomp} 

\usepackage{graphicx}
\usepackage[margin=1.25in]{geometry}
\usepackage[usenames,dvipsnames]{color}

\newcommand\acp{\begin{center}
\rule[-0.2in]{\hsize}{0.01in}\\\rule{\hsize}{0.01in}\\
\vskip 0.1in Submitted to the  Proceedings\\ 
of the African Conference on Fundamental and Applied Physics
    \vskip 0.05in
    {\it Second Edition, ACP2021, March 7--11, 2022 --- Virtual Event}\\
\rule{\hsize}{0.01in}\\\rule[+0.2in]{\hsize}{0.01in} \\
\end{center}}

\usepackage[firstpage=true]{background}
\backgroundsetup{contents={\parbox{6.5in}{\acp}}, scale=1,placement=top,opacity=1,color=black,position={3.25in,1.2in}}

\usepackage{fancyhdr}
\fancypagestyle{plain}{%
  \fancyhf{}%
  \fancyhead[C]{}
  \fancyfoot[C]{\thepage}
}

\fancypagestyle{empty}{%
  \fancyhf{}%
  \fancyhead[C]{{\it ACP2021, March 7--11, 2022 --- Virtual Edition}}
  \fancyfoot[C]{\thepage}
}
\begin{document}

\begin{frontmatter}


\title{ASFAP Working Group Summary of Societal Engagements}

\author[add1]{Mounia Laassiri\corref{cor1}}
\ead{mounia.laassiri@gmail.com}
\author[add2]{Marie Clémentine Nibamureke}
\author[add3]{Bertrand Tchanche Fankam}
\author[add2]{Sam Ramaila}
\author[add2]{Ndeye Arame Boye-Faye}
\author[add4]{Diallo Boye}
\author[add5]{Marie Chantal Cyulinyana}
\author[add6]{Uli Raich}
\author[add7]{Jamal Mimouni}
\author[add8]{Benard Mulilo}
\author[add9]{Iroka Chidinma Joy }

\cortext[cor1]{Corresponding Author}

\address[add1]{Mohamed V University, Morocco}
\address[add2]{University of Johannesburg, South Africa}
\address[add3]{Université Alioune Diop, Senegal}
\address[add4]{Brookhaven National Laboratory, USA}
\address[add5]{University of Rwanda, Rwanda}
\address[add6]{CERN, Switzerland}
\address[add7]{University of Constantine-1 $\&$ CERIST, Algeria}
\address[add8]{University of Zambia, Zambia}
\address[add9]{NASRDA, Nigeria}

\begin{abstract}
\noindent 
The second African Conference of Fundamental and Applied Physics (ACP2021) took place in  the week of March 7–11, 2022. During this conference, all the African Strategy for Fundamental and Applied Physics (ASFAP) working groups had been reserved specials sessions to discuss their scope, activities (past $\&$ current) and topics of common interests. The aim of this report is to summarize the discussion of the ASFAP working groups in societal engagements, namely Physics Education, Community Engagement, Young Physicists and Women in Physics. The recommendations for future activities in societal engagements are summarised in the report as well.
\end{abstract}


\end{frontmatter}

%


\section{Introduction}
\label{sec:intro}
\noindent

The ACP2021~\citep{ACP2021-report, acp2021} was held virtually, with over six hundred and fifty registered participants, five hundred and sixty-three of whom came from thirty-three African countries. The overall \href{https://indico.cern.ch/event/1060503/timetable/#all}{program}, \href{https://indico.cern.ch/event/1060503/contributions/}{presentations} and \href{https://indico.cern.ch/event/1060503/sessions/428245/#all}{recordings} are available at~\cite{acp2021}.\\

The compilation of the ASFAP working group summary of societal engagements is based on discussions and working meetings from the societal engagements community. All major societal engagement projects  in ASFAP have been included in this report and have been categorized as follows:

\begin{itemize}

 \item Societal Engagements: Status $\&$ Plan;
  \item Discussion on Societal Engagements.
\end{itemize}

\section{Societal Engagements: Status $\&$ Plan}
\label{sec:SP}
\noindent

ASFAP contains four working groups devoted to societal engagements. Our task was twofold: to discuss the focus and scope of the working groups in societal engagements; and either identify topics of common interest or topics that are cross-cutting among various societal engagement working groups. 

\subsection{Physics Education}
Research in fundamental and applied physics needs to be supported by a strong and effective physics education to train the next generation of physicists. The objective of the Physics Education (Phys Ed)~\citep{PE} working group is to identify where improvements are needed and propose improved methods to prepare and deliver physics instructions or lessons.\\

As part of the ASFAP process, the Phys Ed working group held workshops, carried out discussions regarding participation and attitudes toward physics education and outreach. Targeted workshops include physics education discussion: at the university level~\citep{PE1W, PE2W}, with the Francophone countries~\citep{PE3W} and with the engagement working groups~\citep{PE4W}. The Phys Ed working group received a number of letter of interests (LOIs)~\cite{LOI} that will be used to form white-paper study groups for the ASFAP strategy.\\

The Phys Ed working group developed its goals, strategies and recommendations from physicists and education input and outreach professionals obtained prior to and during the ACP2021 meeting. The recommendations support a proactive, coordinated Phys Ed working group effort from the entire ASFAP community. The following are recommendations from previous meetings, conferences and workshops:\\

\textbf{Recommendation 1: Increase investment in infrastructure and equipment.}

For Africa to adequately leverage its potential for developing physics education and outreach, a coordinated effort between African countries is needed to ensure that a sustained stream of investment is directed.\\

\textbf{Recommendation 2: Improve the quality of teachings through the training of teachers and a strong focus on applications and technologies.} 

Less theory oriented curricula and more practicals and experimental physics are needed if we want to train good skilled physicists and engineers in African countries. \\

\textbf{Recommendation 3: Pan-African association of physics teachers and lecturers}

We propose to establish a Pan-African association of physics teachers and lecturers. Science societies in Africa should be strongly encouraged to commemorate continental scientific events. 

\subsection{Community Engagement}

Community Engagement (CE)~\citep{CE} working group consists of several sub-groups, namely physics communication and outreach; technology transfer; Internet connectivity/ start-up resources, applications and industry; e-lab $\&$ e-learning; business development and entrepreneurism;  public education and outreach, diversity, inclusion and equity;  government engagement and public policy; and career pipelines $\&$ development, retention and capacity development. The objective of CE working group is to draw a broader engagement and participation in the development of the African strategy, address issues of physics education and intra-African—national, regional and pan-African—collaborations on education and research.\\

The CE working group held five pre-meetings in order to discuss the scope and different topics proposed by the group. February 2021, CE held a workshop with other societal engagement working groups~\citep{CE1W}. At this workshop, discussions led to ideas for future collaborations between CE and other societal engagement working groups. The CE working group has proposed a series of topics in conjunction with other societal engagement working groups. Among them: 

\begin{itemize}
 \item Physics and Environmental Pollution:\\
 How can we use physics to resolve the problem of environmental pollution and raise awareness of the local community on environmental pollution? This would include: recycling methods for plastics, waste burning, special collection programs for pharmaceutical waste, education $\&$ awareness campaigns.
 
  \item Public Outreach $\&$ Education:\\
This will create awareness and broaden the community’s understanding of physics. It would include a survey on the views of physics teachers in Africa, periodic training of physics teachers, annual fairs to introduce the public and in particular the children to the fun of physics. It would also include virtual physics laboratories sessions and campus visits for high school children. Virtual Physics laboratories: for those schools where there is no access to laboratories (+ internet access): classroom demonstrations for teachers and students and physics Olympiads.

  \item Astronomy in the service of physics:\\
  Astronomy is that discipline which can fire up people’s imagination, and hook them to science. It may at the same time play the role of an appetizer for the other sciences in addition to being a fundamental science by itself. Indeed, the Cosmos being after all the largest laboratory in the World. Astronomy could be seen as generalized physics, unless physics (the study of the matter) wishes to reclaim it to itself and considers it as part of it! In this case, physicists will have to get their act together as far as their connection to astronomy.

\end{itemize}

\subsection{Young Physicists Forum}

Young Physicists Forum (YPF)~\citep{YPF} was established to engage rising-star physicists to gather, study and debate the major issues in their research careers. As a guideline, YPF roughly defines young physicist as students, postdocs, engineers, technicians, faculty, etc., up to $\sim$10 years post-highest degree. The main objectives of the YPF are among others, to create a diverse continent of next-generation physicists to play an active role in collaborations pertaining to scientific research and educational issues in Africa. \\

Since launching YPF in 2021, the forum has played an active role in identifying the challenges and remedies for young physicists to flourish in various physics fields. To this effect, the forum has so far conducted several virtual meetings to share the knowledge. January 2022~\citep{YPF1W}, the forum invited stakeholders to discuss some of the challenges and opportunities for young African physicists. The workshop brought together young physicist researchers and feature panelists, drawing more than one hundred and forty registered participants from all over Africa. During presentations and panel discussion, physicists detailed the challenges facing young African physicists; highlighted existing solutions; and brainstormed new strategies for research and policy.\\

\textbf{Physicists Data Collection: ASFAP— Young Physicists Forum Survey}\\

The ASFAP— Young Physicists Forum survey~\citep{Survey} was originally intended to run for five and a half months (from July 15 until December 31, 2021), however, the deadline was extended to December 31, 2022 to allow further feedback from the community. Results from two of the questions are presented below.

\begin{figure}[!htbp]
\begin{center}
\includegraphics[width=\textwidth, height=6cm]{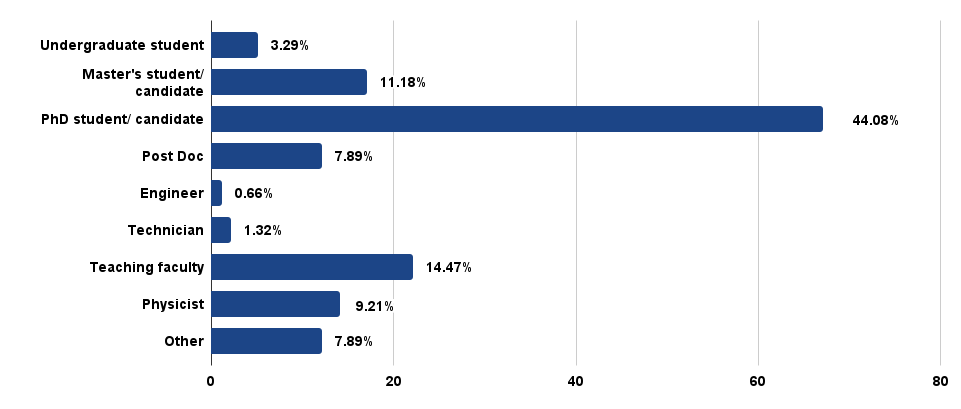}
\end{center}
\caption{Current position of the respondents.}
\label{fig:F1}
\end{figure}

Figure~\ref{fig:F1} shows that majority of the survey respondents are PhD students or candidates (44.08$\%$), followed by teaching faculty members (14.47$\%$) and Master's students or candidates (11.18$\%$).\\

\begin{figure}[!htbp]
\begin{center}
\includegraphics[width=\textwidth, height=6cm]{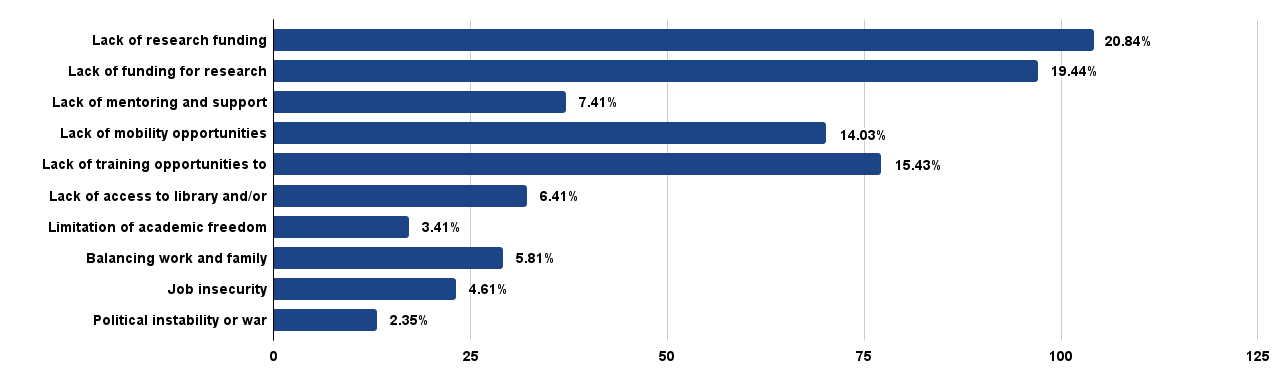}
\end{center}
\caption{Respondents' perceptions of the impact of 10 challenges on their careers.}
\label{fig:F2}
\end{figure}

The results in Figure~\ref{fig:F2} show that general lack of research funding and funding for equipment identified by two-thirds of our respondents as the biggest challenges. Challenges related to human capacity building and professional development (lack of training opportunities; lack of mobility and lack of mentoring and support) were subsequently listed as the third largest challenges by most of our respondents and, importantly, especially by the young physicists. Therefore the need to develop digit libraries supported by efficient Internet connectivity at low costs to African researchers. The fifth largest challenge (balancing work and family demands) speaks to time demands. Interestingly, 'job insecurity' received the lowest rating. Political and social factors (lack of academic freedom and political instability) was also not listed as a major challenge– probably because most of our respondents are already in permanent academic or research positions.\\

The YPF has received a number of LOIs so far and are related to the challenges facing young African physicists~\citep{LOI}. In addition to the LOIs, many YPF topics and issues were collected through workshops and the survey results.

\subsection{Women in Physics Forum}

Despite so many efforts over the past decades to close the gender gap in physics in Africa, women are still largely unrepresented in the physics workforce and even fewer women reach leadership positions.  The objective of Women in Physics Forum (WiPF)~\citep{WiP} is to mitigate the lack of the African women participating in Physics.\\

In the short time since WiPF was launched in 2021, WiPF has started important discussions with ASFAP community regarding women's participation in and attitudes toward physics and raised awareness of the systemic structural barriers that silently push women away from their career track. \\

WiPF organized their first workshop in February 2022~\citep{WiP1W}  which was the international Day for Women and Girls in Science.  On December 9, 2021, WiPF get a chance to participate in The AERAP Africa-Europe Science Collaboration Platform in a session dedicated for women's movements in the fields of science, technology, and innovation. As a result of these many meetings and discussion sessions, WiPF developed the following strategies and implementation plans to help women community achieve the overarching WiPF goals:

\begin{itemize}
 \item Initiate mentorship and networking initiatives.
  \item Establish a system to encourage more girls to study physics.
  \item 
Links to external databases or community-generated databases that track important statistics on women participation in physics.
  \item Emphasize on the gender balance in working place.
  \item Funding of physics projects for women and removing age limit for women in physics.
 \end{itemize}

\section{Discussion on Societal Engagements}
\label{sec:DSE}
\noindent

Following Societal Engagements: Status $\&$ Plan at ACP2021, we had a discussion of a mix of community engagement, physics education, young physicists, and women in physics colleagues. Ideas for future interactions between all societal engagement working groups have been identified.\\

\textbf{Implementation ideas}
\begin{enumerate}
  \item Organise join meetings to disseminate new ideas and work to identify common issues.
  \item Work to identify and improve the the grassroots level related-issues. 
  \item  Create and foster new opportunities for interaction with fields beyond physics.
 \item  Develop and increase access to resources, training activities, and opportunities that engage physicists with policy makers, opinion leaders, the general public, educators and students.

\end{enumerate}

The societal engagements will be engaged in a variety of efforts to inform the public and policy makers about fundamental and applied physics research, and to encourage support for that research. The  societal engagements should augment and enhance ongoing efforts by providing continental coordination and support, and by developing needed resources to make a compelling case for support of fundamental and applied physics research.

\section{Conclusions}
\label{sec:conc}

We summarized past and current activities of the societal engagements community. These include: (1) conferences, workshops, meetings and panel discussions, (2) survey, and (3) letter of interests. These activities have been useful in terms of interest and participation, and should be encouraged and continued. However, we conclude that more direct engagement from all societal engagement working groups is needed.

\section*{Acknowledgments}

Special thanks to the ASFAP— societal engagement co-conveners $\&$  session organizers for their contributions.



\bibliographystyle{elsarticle-num}
\bibliography{myreferences}
\end{document}